\documentclass[prl,preprint]{revtex4}
\usepackage{amsmath}
\usepackage{graphicx}

\begin{document}

\title{Decay of Coherence and Entanglement of a Superposition State for a Continuous Variable System in an Arbitrary Heat Bath}
\author{G. W. Ford}
\affiliation{Department of Physics, University of Michigan, Ann Arbor, MI 48109-1040 USA}
\author{R. F. O'Connell}
\affiliation{Department of Physics and Astronomy, Louisiana State University, Baton
Rouge, LA 70803-4001 USA}
\date{\today}

\begin{abstract}
We consider the case of a pair of particles initially in a superposition
state corresponding to a separated pair of wave packets. We calculate \emph{
exactly} the time development of this non-Gaussian state due to interaction with an \emph{
arbitrary} heat bath. We find that coherence decays continuously, as
expected. We then investigate entanglement and find that at a finite time
the system becomes separable (not entangled). Thus, we see that entanglement sudden death is also prevalent in continuous variable systems which should raise concern for the designers of entangled systems.
\end{abstract}

\maketitle

For continuous variable systems "entanglement sudden death" \cite{yu03},
that is, complete termination of entanglement after a finite time interval,
has been demonstrated for the special case of a pair of particles in a
Gaussian state\cite{dodd041}. Those authors use a master equation and the
necessary and sufficient criterion for separability of such states developed
by Duan et al. \cite{duan00}. Here, we present a more general model by
considering the case of a widely separated pair of particles initially in a
superposition state corresponding to a displaced pair of wave packets. We
use a method that allows us to calculate \emph{exactly} the time development
due to interaction with an \emph{arbitrary }linear passive heat bath \cite
{ford07}. We find first of all that coherence, defined as the relative
amplitude of the interference pattern, decays continuously but very rapidly.
Next we consider entanglement and find that after a finite time the system
becomes separable, showing that \textquotedblleft sudden
death\textquotedblright\ of entanglement occurs for this system as well.

The method is based on the general prescription described in an earlier
publication \cite{ford07} in which a system is put in an initial state by a
measurement applied to the equilibrium state and after a finite time is
sampled by a second measurement. A key formula is the expression for the
Wigner characteristic function given in Eq. (6.5) of \cite{ford07}. For a
two particle system, this formula takes the form:
\begin{equation}
\tilde{W}(Q_{1},P_{1};Q_{2},P_{2})=\frac{\left\langle f^{\dag
}(1)e^{-i(x_{1}(t)P_{1}+m\dot{x}_{1}(t)Q_{1}+x_{2}(t)P_{2}+m\dot{x}
_{2}(t)Q_{2})/\hbar }f(1)\right\rangle }{\left\langle f^{\dag
}(1)f(1)\right\rangle },  \label{1}
\end{equation}
where the initial measurement is described by
\begin{equation}
f(1)=f(x_{1}(0)-x_{1},x_{2}(0)-x_{2}),  \label{2}
\end{equation}
in which $f(x_{1},x_{2})$ is the c-number function describing the initial
measurement while $x_{1}(t)$ and $x_{2}(t)$ are the time-dependent
Heisenberg operators corresponding to the displacement of either particle:
\begin{equation}
x_{j}(t)=e^{iHt/\hbar }x_{j}(0)e^{-iHt/\hbar }.  \label{3}
\end{equation}
Finally, in this formula the brackets indicate expectation with respect to
the state of the system in equilibrium at temperature $T$,
\begin{equation}
\left\langle O\right\rangle =\frac{\mathrm{Tr}\left\{ Oe^{\frac{H}{kT}
}\right\} }{\mathrm{Tr}\left\{ e^{\frac{H}{kT}}\right\} }.  \label{4}
\end{equation}
Here we emphasize that in Eqs. (\ref{3}) and (\ref{4}) $H$ is the
Hamiltonian operator for the \emph{system} of the pair of particles
interacting with the heat bath.

In order to evaluate this formula we make the key assumptions that particles
are linear oscillators coupled to a linear passive heat bath and that within
the bath the particles are \emph{widely separated} so that we may ignore
bath-induced interactions. These assumptions imply that $x_{1}(t)$ and $
x_{2}(t)$ independently undergo quantum Brownian motion. We can now repeat
the discussion leading to Eq. (6.43) of our earlier publication,\cite{ford07}
to obtain
\begin{eqnarray}
&&\left\langle f^{\dag }(1)e^{-i(x_{1}(t)P_{1}+m\dot{x}
_{1}(t)Q_{1}+x_{2}(t)P_{2}+m\dot{x}_{2}(t)Q_{2})/\hbar }f(1)\right\rangle 
\nonumber \\
&=&\exp \{-\sum_{n=1}^{2}\frac{\left\langle x^{2}\right\rangle
(P_{n}^{2}-K_{n}^{2})+m^{2}\left\langle \dot{x}^{2}\right\rangle Q_{n}^{2}}{
2\hbar ^{2}}\}  \nonumber \\
&&\times \int_{-\infty }^{\infty }dx_{1}^{\prime }\int_{-\infty }^{\infty
}dx_{2}^{\prime }f^{\dag }(x_{1}^{\prime }+\frac{L_{1}}{2},x_{2}^{\prime }+
\frac{L_{2}}{2})f(x_{1}^{\prime }-\frac{L_{1}}{2},x_{2}^{\prime }-\frac{L_{2}
}{2})  \nonumber \\
&&\times \frac{1}{2\pi \left\langle x^{2}\right\rangle }\exp
\{-\sum_{n=1}^{2}\frac{(x_{n}+x_{n}^{\prime })^{2}}{2\left\langle
x^{2}\right\rangle }-i(x_{n}+x_{n}^{\prime })\frac{K_{n}}{\hbar }\},
\label{5}
\end{eqnarray}
where $\left\langle x^{2}\right\rangle $ and $\left\langle \dot{x}
^{2}\right\rangle $ are the mean squares of the displacement and velocity,
the same for either particle, and \ we have introduced
\begin{equation}
K_{n}=\frac{cP_{n}+m\dot{c}Q_{n}}{\left\langle x^{2}\right\rangle },\quad
L_{n}=GP_{n}+m\dot{G}Q_{n}.  \label{6}
\end{equation}
Here $G=G(t)$ is the Green function and $c=c(t)\equiv \frac{1}{2}\langle
x(t)x(0)+x(0)x(t)\rangle $ is the correlation function, again the same for
either particle.

These expressions are valid for any measurement function. We now specialize
to the case of a pair of particles initially in a superposition state
corresponding to a separated pair of wave packets, with measurement function
of the form:
\begin{eqnarray}
f(x_{1},x_{2}) &=&\frac{1}{\sqrt{4\pi \sigma ^{2}(1+e^{-d^{2}/4\sigma ^{2}})}
}\left[ \exp \left\{ -\frac{(x_{1}-d/2)^{2}+(x_{2}+d/2)^{2}}{4\sigma ^{2}}
\right\} \right.  \nonumber \\
&&+\left. \exp \left\{ -\frac{(x_{1}+d/2)^{2}+(x_{2}-d/2)^{2}}{4\sigma ^{2}}
\right\} \right] .  \label{7}
\end{eqnarray}
Here we emphasize that the wave packet separation $d$ is arbitrary and
should not be confused with the separation of the particles in the bath,
which is large.

With this measurement function the integrals in the expression (\ref{5}) are
standard Gaussian \cite{ford07}. Putting the result in the expression (\ref
{1}) for the Wigner characteristic function we find
\begin{eqnarray}
\tilde{W}(Q_{1},P_{1};Q_{2},P_{2};t)  \nonumber \\
=\exp \left\{ -\sum_{n=1}^{2}\frac{\left\langle x^{2}\right\rangle \left(
P_{n}^{2}-\frac{\left\langle x^{2}\right\rangle K_{n}^{2}}{\left\langle
x^{2}\right\rangle +\sigma ^{2}}\right) +m^{2}\left\langle \dot{x}
^{2}\right\rangle Q_{n}^{2}+\frac{\hbar ^{2}}{4\sigma ^{2}}L_{n}^{2}}{2\hbar
^{2}}\right\}  \nonumber \\
\times \frac{\cos \frac{\left\langle x^{2}\right\rangle (K_{1}-K_{2})d}{
2\hbar (\left\langle x^{2}\right\rangle +\sigma ^{2})}+\exp \left\{ -\frac{
\left\langle x^{2}\right\rangle d^{2}}{4\sigma ^{2}(\left\langle
x^{2}\right\rangle +\sigma ^{2})}\right\} \cosh \left\{ \frac{(L_{1}-L_{2})d
}{4\sigma ^{2}}\right\} }{1+\exp \left\{ -\frac{\left\langle
x^{2}\right\rangle d^{2}}{4\sigma ^{2}(\left\langle x^{2}\right\rangle
+\sigma ^{2})}\right\} }.  \label{8}
\end{eqnarray}
where in order to center the state at the origin we have put $x_{1}=x_{2}=0$.

This\ expression becomes simpler in the free particle limit :$\ \left\langle
x^{2}\right\rangle \rightarrow \infty $. In this limit
\begin{eqnarray}
&&\tilde{W}(Q_{1},P_{1};Q_{2},P_{2};t)  \nonumber \\
&=&\exp \left\{ -\frac{
A_{11}(P_{1}^{2}+P_{2}^{2})+2A_{12}(Q_{1}P_{1}+Q_{2}P_{2})+A_{22}(Q_{1}^{2}+Q_{2}^{2})
}{2\hbar ^{2}}\right\}  \nonumber \\
&&\times \frac{\cos \frac{(P_{1}-P_{2})d}{2\hbar }+\exp \left\{ -\frac{d^{2}
}{4\sigma ^{2}}\right\} \cosh \left\{ \frac{\left[ G\left(
P_{1}-P_{2}\right) +m\dot{G}\left( Q_{1}-Q_{2}\right) \right] d}{4\sigma ^{2}
}\right\} }{1+\exp \left\{ -\frac{d^{2}}{4\sigma ^{2}}\right\} },  \label{9}
\end{eqnarray}
in which we have introduced
\begin{eqnarray}
A_{11} &=&\sigma ^{2}+s+\frac{\hbar ^{2}G^{2}}{4\sigma ^{2}},  \nonumber \\
A_{12} &=&\frac{m\dot{s}}{2}+\frac{\hbar ^{2}m\dot{G}G}{4\sigma ^{2}}, 
\nonumber \\
A_{22} &=&m^{2}\left\langle \dot{x}^{2}\right\rangle +\frac{\hbar ^{2}m^{2}
\dot{G}^{2}}{4\sigma ^{2}}.  \label{10}
\end{eqnarray}
In these expressions $s=2\left( \left\langle x^{2}\right\rangle -c\right)
=\left\langle \left( x(t)-x(0)\right) ^{2}\right\rangle $ is the mean square
displacement and as above $G$ is the Green function.

The Wigner function is the inverse Fourier transform of the Wigner
characteristic function:
\begin{eqnarray}
W(q_{1},p_{1};q_{2},p_{2};t) &=&\frac{1}{2(1-e^{-d^{2}/4\sigma ^{2}})}\left[
W_{0}(q_{1}-\frac{d}{2},p_{1};t)W_{0}(q_{2}+\frac{d}{2},p_{2};t)\right. 
\nonumber \\
&&+W_{0}(q_{1}+\frac{d}{2},p_{1};t)W_{0}(q_{2}-\frac{d}{2},p_{2};t)  \nonumber
\\
&&\left. +2e^{-A(t)}W_{0}(q_{1},p_{1};t)W_{0}(q_{2},p_{2};t)\cos \Phi
(q_{1}-q_{2},p_{1}-p_{2};t)\right] .  \label{11}
\end{eqnarray}
Here $W_{0}$ is the Wigner function for a single particle wave packet,
\begin{equation}
W_{0}(q,p;t)=\frac{1}{2\pi \sqrt{A_{11}A_{22}-A_{12}^{2}}}\exp \left\{ 
\mathbf{-}\frac{A_{22}q^{2}-2A_{12}qp+A_{11}p^{2}}{2(A_{11}A_{22}-A_{12}^{2})
}\right\} ,  \label{12}
\end{equation}
while the phase $\Phi $ is given by
\begin{equation}
\Phi (q,p;t)=\frac{(GA_{22}-m\dot{G}A_{12})q+(m\dot{G}A_{11}-GA_{12})p}{
A_{11}A_{22}-A_{12}^{2}}\frac{\hbar d}{4\sigma ^{2}}  \label{13}
\end{equation}
and the quantity $A$ by
\begin{equation}
A(t)=\frac{(A_{11}-\frac{\hbar ^{2}G^{2}}{4\sigma ^{2}})(A_{22}-\frac{\hbar
^{2}m^{2}\dot{G}^{2}}{4\sigma ^{2}})-(A_{12}-\frac{\hbar ^{2}mG\dot{G}}{
4\sigma ^{2}})^{2}}{A_{11}A_{22}-A_{12}^{2}}\frac{d^{2}}{4\sigma ^{2}}.
\label{14}
\end{equation}
We note that each of the first two terms in brackets in the expression (\ref
{8}) for the Wigner function corresponds to the product of independently
propagating packets. We call these the direct terms. The third term is an
interference term. We emphasize that we have assumed that the particles are
widely separated within the bath so there is no coupling between them. The
presence of this interference term is therefore a purely quantum mechanical
phenomenon.

The Wigner function is a quasiprobability distribution, not directly
observable. A physical observable is the probability distribution, obtained
by integrating over the momentum variables:
\begin{eqnarray}
P(q_{1},q_{2};t) &=&\frac{1}{2(1-e^{-d^{2}/4\sigma ^{2}})}\left[ P_{0}(q_{1}-
\frac{d}{2},t)P_{0}(q_{2}+\frac{d}{2},t)\right.  \nonumber \\
&&+P_{0}(q_{1}+\frac{d}{2},t)P_{0}(q_{2}-\frac{d}{2},t)  \nonumber \\
&&\left. +2a(t)\exp \left\{ -\frac{d^{2}}{4A_{11}}\right\}
P_{0}(q_{1};t)P_{0}(q_{2};t)\cos \left\{ \frac{\hbar Gd\left(
q_{1}-q_{2}\right) }{4A_{11}\sigma ^{2}}\right\} \right] .  \label{15}
\end{eqnarray}
Again, the first two terms are direct terms corresponding to independently
propagating wave packets with
\begin{equation}
P_{0}(q;t)=\frac{1}{\sqrt{2\pi A_{11}}}\exp \left\{ -\frac{q^{2}}{2A_{11}}
\right\} ,  \label{16}
\end{equation}
the probability distribution for a single wave packet centered at the
origin. The third term is an interference term. Viewed in the $\left(
q_{1},q_{2}\right) $ plane, the direct terms are seen as a pair of peaks
centered at $\left( q_{1},q_{2}\right) =\left( \frac{d}{2},-\frac{d}{2}
\right) $ and $\left( q_{1},q_{2}\right) =\left( -\frac{d}{2},\frac{d}{2}
\right) $ and spreading in time as the width $A_{11}$ increases. The
interference term is seen as a spreading peak centered at the origin \ and
modulated by the cosine term. The quantity $a(t)$ is the ratio of the
geometric mean of the direct term to the factor multiplying the cosine in
the interference term and is therefore a measure of the visibility of the
interference. We find
\begin{equation}
a(t)=\exp \left\{ -\frac{s(t)}{\sigma ^{2}+s(t)+\frac{\hbar ^{2}G^{2}(t)}{
4\sigma ^{2}}}\frac{d^{2}}{4\sigma ^{2}}\right\} .  \label{17}
\end{equation}
This quantity is initially unity and, for $d$ large, diminishes rapidly to a
very small asymptotic value. This is the familiar phenomenon of decoherence
of a superposition state. But nevertheless interference is present for all
times, albeit with a small amplitude. Our point here is that there is no
sudden death of coherence as indicated by the presence of the interference
term.

We turn now to the question of entanglement. A two-particle state described
by a density matrix $\rho $ is said to be separable (not entangled) if and
only if $\rho $ can expressed in the form
\begin{equation}
\rho =\sum_{j}p_{j}\rho _{j}(1)\rho _{j}(2),  \label{18}
\end{equation}
in which $\rho _{j}(1)$ and $\rho _{j}(2)$ are projection operators into
states of particles $1$ and $2$, respectively, and the $p_{j}$ are positive.
In our case we seek to express the density matrix elements in the form
\begin{equation}
\left\langle x_{1}^{\prime },x_{2}^{\prime }\left\vert \rho \right\vert
x_{1},x_{2}\right\rangle =\int d^{2}\alpha _{1}\int d^{2}\alpha _{2}P\left(
\alpha _{1},\alpha _{2}\right) \phi _{\alpha _{1}}(x_{1}^{\prime })\phi
_{\alpha _{1}}^{\ast }(x_{1})\phi _{\alpha _{2}}(x_{2}^{\prime })\phi
_{\alpha _{2}}^{\ast }(x_{2}),  \label{19}
\end{equation}
where the $\phi $'s are what we might call strong form coherent wave
functions:
\begin{equation}
\phi _{\alpha }\left( x\right) =\left( 2\pi \sigma _{0}^{2}\right)
^{-1/4}\exp \left\{ -\frac{1-i\delta _{0}}{4\sigma _{0}^{2}}\left( x-\bar{x}
\right) ^{2}+\frac{i\bar{p}x}{\hbar }-i\frac{\bar{x}\bar{p}}{2\hbar }
\right\} ,  \label{20}
\end{equation}
with the state labelled with the complex number
\begin{equation}
\alpha =\frac{1-i\delta _{0}}{2\sigma _{0}}\bar{x}+i\frac{\sigma _{0}}{\hbar 
}\bar{p},\quad d^{2}\alpha =\frac{d\bar{x}d\bar{p}}{2\hbar }.  \label{21}
\end{equation}
This is clearly of the form (\ref{18}) with the sum replaced by an integral,
so if this expansion exists and $P\left( \alpha _{1},\alpha _{2}\right) $ is
everywhere positve the state is separable. The expression (\ref{19}) is
reminiscent of the Glauber-Sudarshan $P$-representation, \cite{hillery84}
but in that representation the $\phi $'s are coherent states, which are
expressed in terms of the ground state of an oscillator, or equivalently a
minimum uncertainty state, \cite{ford02} shifted in position and momentum.
If in the wave function (\ref{20}) we set the parameter $\delta _{0}$ equal
to zero we have such a coherent wave function. On the other hand, if $\delta
_{0}$ is not zero, the wave function (\ref{20}) minimizes the strong form of
the uncertainty relation: \cite{schrodinger30,sakurai85}
\begin{equation}
\left\langle \left( x-\bar{x}\right) ^{2}\right\rangle \left\langle \left( p-
\bar{p}\right) ^{2}\right\rangle -\left\langle \frac{\left( x-\bar{x}\right)
\left( p-\bar{p}\right) +\left( p-\bar{p}\right) \left( x-\bar{x}\right) }{2}
\right\rangle ^{2}\geq \frac{\hbar ^{2}}{4}.  \label{22}
\end{equation}
It is not difficult to show that wave function (\ref{20}) satisfies this as
an equality.

Next, we recall the relation between the Wigner characteristic function and
the density function matrix elements:
\begin{equation}
\tilde{W}(Q_{1},P_{1};Q_{2},P_{2})=\int dq_{1}\int
dq_{2}e^{-i(q_{1}P_{1}+q_{2}P_{2})/\hbar }\left\langle q_{1}-\frac{Q_{1}}{2}
,q_{2}-\frac{Q_{2}}{2}\left\vert \rho \right\vert q_{1}+\frac{Q_{1}}{2}
,q_{2}+\frac{Q_{2}}{2}\right\rangle .  \label{23}
\end{equation}
Using the expansion (\ref{19}) of the density matrix elements, this becomes
\begin{eqnarray}
\tilde{W}(Q_{1},P_{1};Q_{2},P_{2}) &=&\int d^{2}\alpha _{1}\int d^{2}\alpha
_{2}P\left( \alpha _{1},\alpha _{2}\right) \int dq_{1}\phi _{\alpha
_{1}}(q_{1}-\frac{Q_{1}}{2})\phi _{\alpha _{1}}^{\ast }(q_{1}+\frac{Q_{1}}{2}
)e^{-iq_{1}P_{1}/\hbar }  \nonumber \\
&&\times \int dq_{2}\phi _{\alpha _{2}}(q_{2}-\frac{Q_{2}}{2})\phi _{\alpha
_{2}}^{\ast }(q_{2}+\frac{Q_{2}}{2})e^{-iq_{2}P_{2}/\hbar }.  \label{24}
\end{eqnarray}
With the explicit form (\ref{20}) of the coherent state, we see that
\begin{equation}
\int dq\phi _{\alpha }(q-\frac{Q}{2})\phi _{\alpha }^{\ast }(q+\frac{Q}{2}
)e^{-iqP/\hbar }=e^{-i\frac{\bar{p}Q+\bar{x}P}{\hbar }}\exp \left\{ -\frac{
\sigma _{0}^{2}P^{2}}{2\hbar ^{2}}-\frac{\delta _{0}QP}{2\hbar }-\frac{
\left( 1+\delta _{0}^{2}\right) Q^{2}}{8\sigma _{0}^{2}}\right\} .
\label{25}
\end{equation}
Therefore, the expression (\ref{24}) can be written
\begin{eqnarray}
&&\int d^{2}\alpha _{1}\int d^{2}\alpha _{2}P\left( \alpha _{1},\alpha
_{2}\right) e^{-i\left( \bar{p}_{1}Q_{1}+\bar{x}_{1}P_{1}+\bar{p}_{2}Q_{2}+
\bar{x}_{2}P_{2}\right) /\hbar }  \nonumber \\
&=&\exp \left\{ \sum_{j=1,2}\frac{\sigma _{0}^{2}P_{j}^{2}+\hbar \delta
_{0}Q_{j}P_{j}+\frac{\hbar ^{2}\left( 1+\delta _{0}^{2}\right) }{4\sigma
_{0}^{2}}Q_{j}^{2}}{2\hbar ^{2}}\right\} \tilde{W}(Q_{1},P_{1};Q_{2},P_{2}).
\label{26}
\end{eqnarray}
This is just the Fourier transform of the $P$-function, which will exist if
the inverse transform exists. From an inspection of the Wigner
characteristic function (\ref{9}) for our superposition state, we see that
convergence of the inverse transform will be dominated by the exponential
factors and will therefore exist if the quadratic form
\begin{equation}
\left( 
\begin{array}{cc}
P & Q
\end{array}
\right) \left( 
\begin{array}{cc}
A_{11}-\sigma _{0}^{2} & A_{12}-\frac{\hbar \delta _{0}}{2} \\ 
A_{12}-\frac{\hbar \delta _{0}}{2} & A_{22}-\frac{\hbar ^{2}\left( 1+\delta
_{0}^{2}\right) }{4\sigma _{0}^{2}}
\end{array}
\right) \left( 
\begin{array}{c}
P \\ 
Q
\end{array}
\right)  \label{27}
\end{equation}
is positive definite. Since the parameters $\sigma _{0}$ and $\delta _{0}$
are arbitrary we can first choose $\delta _{0}$ to make this quadratic form
diagonal and then choose $\sigma _{0}$ to minimize the product of the
diagonal elements. The corresponding optimum values are
\begin{equation}
\left( \delta _{0}\right) _{\text{opt}}=\frac{2A_{12}}{\hbar },\quad \left(
\sigma _{0}^{2}\right) _{\text{opt}}=\sqrt{\frac{\left( \hbar
^{2}+4A_{12}^{2}\right) A_{11}}{4A_{22}}}.  \label{28}
\end{equation}
With this choice we find for the diagonal elements that the diagonal
elements of the quadratic form (\ref{27}) are given by
\begin{eqnarray}
\tilde{A}_{11} &=&\sqrt{\frac{A_{11}}{A_{22}}}\left( \sqrt{A_{11}A_{22}}-
\sqrt{A_{12}^{2}+\frac{\hbar ^{2}}{4}}\right) ,  \nonumber \\
\tilde{A}_{22} &=&\sqrt{\frac{A_{22}}{A_{11}}}\left( \sqrt{A_{11}A_{22}}-
\sqrt{A_{12}^{2}+\frac{\hbar ^{2}}{4}}\right) .  \label{29}
\end{eqnarray}
It is not difficult to see that these are positive at all times. Thus the
expansion (\ref{19}) exists at all times.

Next we consider the positivity of $P\left( \alpha _{1},\alpha _{2}\right) $
. With the opimum values (\ref{28}) of the parameters in (\ref{26}) we form
the inverse Fourier transform. The integrals are all standard Gaussian and
the result can be written in the form
\begin{eqnarray}
P\left( \alpha _{1},\alpha _{2}\right) =\frac{\hbar ^{2}\exp \left\{ -\frac{
\bar{p}_{1}^{2}+\bar{p}_{2}^{2}}{2\tilde{A}_{22}}-\frac{\bar{x}_{1}^{2}+\bar{
x}_{2}^{2}}{4\tilde{A}_{11}}-\left( 1-\frac{\hbar ^{2}G^{2}}{4\sigma ^{2}
\tilde{A}_{11}}-\frac{\hbar ^{2}\sigma ^{2}m^{2}\dot{G}^{2}}{4\tilde{A}_{22}}
\right) \frac{d^{2}}{4\sigma ^{2}}\right\} }{\pi ^{2}\tilde{A}_{11}\tilde{A}
_{22}\left( 1+\exp \left\{ -\frac{d^{2}}{4\sigma ^{2}}\right\} \right) } 
\nonumber \\
\left[ \exp \left\{ \left( 1-\frac{\hbar ^{2}G^{2}}{4\sigma ^{2}\tilde{A}
_{11}}-\frac{\hbar ^{2}\sigma ^{2}m^{2}\dot{G}^{2}}{4\tilde{A}_{22}}-\frac{
\sigma ^{2}}{\tilde{A}_{11}}\right) \frac{d^{2}}{4\sigma ^{2}}\right\} \cosh 
\frac{\left( \bar{x}_{1}-\bar{x}_{2}\right) d}{2\tilde{A}_{11}}\right. 
\nonumber \\
\left. +\cos \left( \frac{\hbar Gd\left( \bar{x}_{1}-\bar{x}_{2}\right) }{
4\sigma ^{2}\tilde{A}_{11}}+\frac{\hbar m\dot{G}d\left( \bar{p}_{1}-\bar{p}
_{2}\right) }{4\sigma ^{2}\tilde{A}_{22}}\right) \right] .  \label{30}
\end{eqnarray}
The first line in Eq. (\ref{30}) is a positive factor, so $P\left( \alpha
_{1},\alpha _{2}\right) $ is positive if the remaining factor is positive.
Clearly this will be the case for all $\left( \bar{x}_{1},\bar{p}_{1},\bar{x}
_{2},\bar{p}_{2}\right) $ if and only if
\begin{equation}
C(t)\equiv 1-\frac{\hbar ^{2}G^{2}}{4\sigma ^{2}\tilde{A}_{11}}-\frac{\hbar
^{2}m^{2}\dot{G}^{2}}{4\sigma ^{2}\tilde{A}_{22}}-\frac{\sigma ^{2}}{\tilde{A
}_{11}}>0.  \label{31}
\end{equation}

At short times $G(t)\cong t/m$ and $s(t)\cong \left\langle \dot{x}
^{2}\right\rangle t^{2}$. With the expressions (\ref{10}) for $A_{11\text{,}
} $ $A_{12}$ and $A_{22}$ and these in turn in the expressions (\ref{29})
for $\tilde{A}_{11}$ and $\tilde{A}_{22}$ we find
\begin{equation}
C(0)=-\frac{\sqrt{U}+1}{U-\sqrt{U}}=-\frac{1+\left( 1+4\sigma ^{2}/\bar{
\lambda}^{2}\right) ^{-1/2}}{\left( 1+4\sigma ^{2}/\bar{\lambda}^{2}\right)
^{1/2}-1},  \label{32}
\end{equation}
where $\bar{\lambda}=\hbar /m\sqrt{\left\langle \dot{x}^{2}\right\rangle }$
is the deBroglie wavelength. Not surprisingly $C(0)$ is always negative,
since the initial state is formed with a projection operator (\ref{7})
corresponding to a necessarily entangled state.

At very long times, the behavior of $G(t)$ and $s(t)$ depends upon the bath
parameters.\cite{ford06} As an illustration we consider the Ohmic model for
\ which at long times
\begin{eqnarray}
G(t) &\sim &\zeta ^{-1},  \nonumber \\
s(t) &\sim &\frac{2\hbar }{\pi \zeta }\log \frac{\zeta t}{m},\qquad T=0, 
\nonumber \\
s(t) &\sim &\frac{2kT}{\zeta }t,\qquad T>0,  \label{33}
\end{eqnarray}
where $\zeta $ is the Ohmic friction constant. With this it is easy to see
that for this Ohmic case at long times $C(t)\sim 1$. Clearly there must be
an intermediate time at which $C(t)$ changes sign and the state becomes
separable. For example, in Fig. 1 we plot $C(t)$ versus $\gamma t$ for the single relaxation time ($\tau$) model \cite{ford07} at zero temperature, where $\tau =\gamma^{-1}/6$ and $\gamma$ is the Ohmic relaxation time. There we see that the change of sign occurs at $\gamma t\approx 6$.  In general, most other bath models (that is, models with colored noise \cite{ford06}) show similar behavior. 

The work of R. F. O'Connell was supported in part by the National Science
Foundation under Grant No. ECCS-0757204.

\begin{figure}
\includegraphics{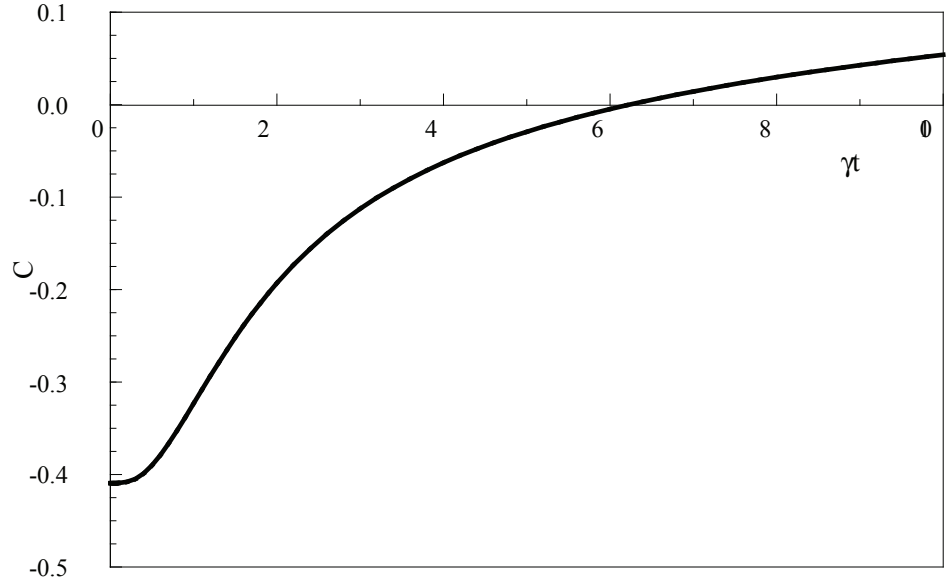}
\caption{$C(t)$ versus $\gamma t$ for the single relaxation time $(\tau )$ model at zero temperature, where $\tau =\gamma^{-1}/6$ and $\gamma $ is the Ohmic relaxation time. We note that separability occurs for $\gamma t \gtrsim 6$.}
\end{figure}

\end{document}